\documentclass[twocolumn,showpacs,aps,floatfix]{revtex4}
\usepackage{amsmath}
\usepackage{graphicx}
\usepackage{dcolumn}
\usepackage{bm}

\begin{document}

\title{Theoretical energies and
transition probabilities of argon}
\author{I. M. Savukov}
 \email{isavukov@princeton.edu}
 \homepage{http://www.princeton.edu/~isavukov}

\affiliation{Department of Physics, Princeton University,
Princeton, New Jersey 08544}

\date{\today}
\begin{abstract}
Argon energies of J=1 odd and J=2 even states, oscillator
strengths of the two lowest J=1 odd states, and line strengths for
transitions between J=1 odd and J=2 even excited states  are
calculated with the CI+MBPT method. The results are in good
agreement with experiment.
\end{abstract}

\maketitle

\section{Introduction}

Given that a combined configuration-interaction (CI) and
many-body-perturbation-theory (MBPT) approach could describe well
experimental data for energies and transition rates in neon, it
was a natural continuation to investigate a more difficult,
stronger core-correlated atom such as argon. In argon theories and
experiments appear to be less accurate than in neon, with larger
mutual disagreement; moreover, heavier closed-shell atoms are even
less studied and understood so that the argon atom can serve as an
intermediate step in understanding many closed-shell atoms and
ions.

Although semiempirical
calculations~\cite{Lilly,gvb,Wende,Bues,jones} sometimes lead to a
good agreement with experiment for some transition arrays and
provide valuable information for the analysis of existing
experimental data, they do not explain much the nature of
correlations and their predictive power is limited to cases  where
(a) energies are well-known and (b) energy-level information is
sufficient to infer other atomic properties such as transition
probabilities. In general two theories could give exactly the same
energy levels without predicting the same transition
probabilities. For example, adding or not adding random-phase
approximation (RPA) corrections to transition amplitudes will give
two different answers for transition rates although wave functions
obtained from the same effective Hamiltonian are identical. When
RPA corrections were included in neon~\cite{isavneonepr}, the
improvement of agreement with experiment was only marginal and it
was not an issue, but as we will show in this paper this type
correction is crucial in argon. Hence, theories, which do not take
into account at least the RPA correction, can be misleading in
argon and heavier noble gas atoms. Still owing to cancellations of
RPA contributions for the same transition array, ratios obtained
with a matrix method could be still accurate. Indeed, a fair
agreement with experiment for low-lying argon transitions is
obtained by~\citet{Lilly} even without considering RPA
corrections. However, the accuracy of ratios decreases if
relativistic and correlation effects, that introduce asymmetry for
the p$_{3/2}$ and p$_{1/2}$ hole RPA contributions to radial
transition integrals, are significant.

Another issue is the role of configuration interaction since many
theories are limited by considering only a few configurations. The
convergence of CI has to be investigated. Furthermore, other
many-body effects, such as an interaction of a valence electron
with a core or a hole with a core or their combination, are quite
significant in argon and heavier noble-gas atoms. Thus, the
development of {\it ab initio} theory for heavy noble-gas atoms is
needed to single out important correlation and relativistic
effects. Some success is already achieved in this direction. For
example, elaborate {\it ab initio} calculations by~\citet{neon:98}
performed for transitions from the lowest odd excited states
$\left[ 3p_{3/2}^{-1}4s_{1/2}\right] _{1}$ and $\left[
3p_{1/2}^{-1}4s_{1/2}\right] _{1}$ to the ground state give
oscillator strengths in good agreement with average experimental
values. However, this theory was not extended to transitions
between excited states. Therefore, one objective of this paper is
to prove that transitions between excited states of argon can also
be calculated with {\it ab initio} methods. We also would like to
show that for argon our simpler CI+MBPT theory has comparable
accuracy to complicated single-double couple-cluster (SDCC) method
of Ref.~\cite{neon:98} and thus can be of practical value for
other theoretical groups.

In this paper, first we will briefly describe our method of
calculations (more details are given in~\cite{neonepr,disser});
then, we will calculate energies and oscillator strengths of J=1
odd states and  energies of J=2 even states and compare them with
experiment. This comparison gives an estimate on accuracy of our
wave functions. Next, we will present line strengths and rates for
transitions between excited states and compare our result with
experiment and semiempirical calculations.

\section{CI+MBPT method}
We have developed a convergent variant of CI+MBPT for
particle-hole states of closed-shell atoms~\cite{disser} and
successfully applied it in neon to calculate
energies~\cite{neonepr} and oscillator strengths for transitions
between excited states~\cite{isavneonepr}. This perturbation
theory is based on the first iteration of couple-cluster
single-double equations~\cite{kaldor,neon:95}. All second-order
MBPT terms~\cite{neon:01} are included, with some denominators
being modified to take into account the strong interaction between
a hole and a core electron or core and core electrons
non-perurbatively~\cite{disser,neonepr}. The advantage of this
approach compared to the iterated single-double
method~\cite{neon:95} is simplicity and speed of calculations. The
accuracy of hole energies and fine-structure splittings is
significantly improved already after adding second-order MBPT
corrections as was illustrated previously in neon. Apart from
Coulomb correlation corrections, the Breit magnetic interaction is
also included, but small frequency-dependent Breit,
quantum-electrodynamic, reduced-mass, and mass-polarization
corrections are omitted.

To calculate particle-hole energies, we construct a model CI
space, compute effective Hamiltonian in this space, and solve an
eigenvalue problem. Along with energies we obtain wave functions,
which are used to calculate transition amplitudes or other related
quantities.  We investigate energies, oscillator strengths for
transitions to the ground state, and line strengths or transition
rates for transitions between excited states. Energies of neon
particle-hole J=1 odd states and oscillator strengths were in very
good agreement with experiment: pure \textit{ab-initio} neon
energies differed from experimental energies by 0.0069 a.u., but
after subtraction of the systematic shift (which does not make
much difference in transition calculations), the agreement was at
the level of 0.0001 a.u.  for almost all states. In this paper we
use the same approach to calculate energies and transition
amplitudes of argon.

Reduced matrix elements for transitions between particle-hole
excited states are given in Ref.~\cite{isavneonepr}:
\begin{multline}
\left\langle F\left\| Z_{J}\right\| I\right\rangle =\sqrt{\left(
2J_{F}+1\right) \left( 2J_{I}+1\right) }  \nonumber \\
\left\{ (-1)^{J+J_{I}+j_{a}+j_{v^{\prime }}}\left\{
\begin{array}{lll}
J_{I} & J & J_{F} \\
j_{v^{\prime }} & j_{a} & j_{v}
\end{array}
\right\} \delta _{a^{\prime }a}\left\langle v^{\prime }\left\|
Z_{J}\right\|
v\right\rangle +\right.   \nonumber \\
\left. (-1)^{J_{F}+j_{a^{\prime }}+j_{v^{\prime }}+1}\left\{
\begin{array}{lll}
J & J_{I} & J_{F} \\
j_{v} & j_{a^{\prime }} & j_{a}
\end{array}
\right\} \delta _{v^{\prime }v}\left\langle a\left\| Z_{J}\right\|
a^{\prime }\right\rangle \right\}
\end{multline}
We found that second-order corrections to transition amplitudes or
lowest order RPA corrections
\begin{equation}
Z_{ij}^{\rm RPA}=Z_{ij}+\sum\limits_{bm}\frac{Z_{bm}\,
\tilde{g}_{imjb}} {\epsilon _{b}-\epsilon _{m}-\omega
}+\sum\limits_{bm} \frac{\tilde{g}_{ibjm}\, Z_{mb}}{\epsilon
_{b}-\epsilon _{m}+\omega }\label{RPA1}
\end{equation}
 are significant in argon. This type
correction requires careful analysis since the possible
interaction between an ``observing" hole state $a$ and core states
$b$ over which summation is carried out should be taken into
account in all orders, at least a bulk part of this interaction.
Similar to our approach for energies, we include the interaction
in all orders by adding the extra term
\begin{equation}
  extra=-\frac{X_0(abab)}{\sqrt{(2j_a+1)(2j_b+1)}}\label{extra}
\end{equation}
in the denominators of RPA expressions. This procedure reduces RPA
corrections roughly twice. If we note that the value of RPA
correction is comparable to the values of lowest order matrix
elements, the modification of the denominators is important and
brings our theory to a good agreement with experiment which will
be illustrated later in this paper.

\section{Calculations for argon}

\subsection{Energies of argon odd J=1 states}

Table~\ref{arg21} shows our calculated energies for J=1 odd
states. Fine structure for the  4s, 5s, or 4f groups of states is
well reproduced since the deviation from experiment for each fine
structure group of levels is similar. The last column shows
shifted deviations from experiment, which demonstrate that there
is a systematic shift for energy levels in our calculations. This
shift is due to the inaccuracy of our calculations of hole
energies. The shifted deviations illustrate a good precision for
higher excited states which have smaller correlation effects. We
also show the convergence with the number of particle-hole
configurations. The change in energy is significant when we
increase the size of CI space from 22 to 32 and to 52; however,
from 52 to 62 the change is small, and we consider that CI-52 case
is optimal.

\begin{table}[tbp]
\caption{Energies of argon odd J=1 states. For unique
specification of levels, accurate NIST energies are provided. The
sizes of CI model spaces are denoted CI-22, etc. $\Delta_{7}$ is
the deviation from experiment for the 7th level. All energies are
expressed in atomic units} \label{arg21}
\begin{center}
\begin{tabular}{ccccccc}
\hline\hline State & NIST  & CI-22  & CI-32  & CI-52  & $\Delta $
& $\Delta-\Delta_{7}$ \\ \hline
p$_{3/2}^{-1}$4s & 0.4272 & 0.3202 & 0.4219 & 0.4219 & 0.0053 & 0.0129 \\
p$_{1/2}^{-1}$4s & 0.4347 & 0.4219 & 0.4291 & 0.4297 & 0.0050 & 0.0126 \\
p$_{3/2}^{-1}$3d & 0.5095 & 0.4295 & 0.4348 & 0.5083 & 0.0012 & 0.0088 \\
p$_{3/2}^{-1}$5s & 0.5178 & 0.5082 & 0.5085 & 0.5250 & -0.0072 & 0.0004 \\
p$_{3/2}^{-1}$3d & 0.5201 & 0.5234 & 0.5257 & 0.5268 & -0.0067 & 0.0009 \\
p$_{1/2}^{-1}$5s & 0.5239 & 0.5268 & 0.5268 & 0.5317 & -0.0078 & -0.0002 \\
p$_{1/2}^{-1}$3d & 0.5256 & 0.5331 & 0.5330 & 0.5332 & -0.0076 & 0.0000 \\
 \hline
\end{tabular}
\end{center}
\end{table}
\subsection{Oscillator strengths of resonant transitions}
\begin{figure}
\centerline{\includegraphics*[scale=0.75]{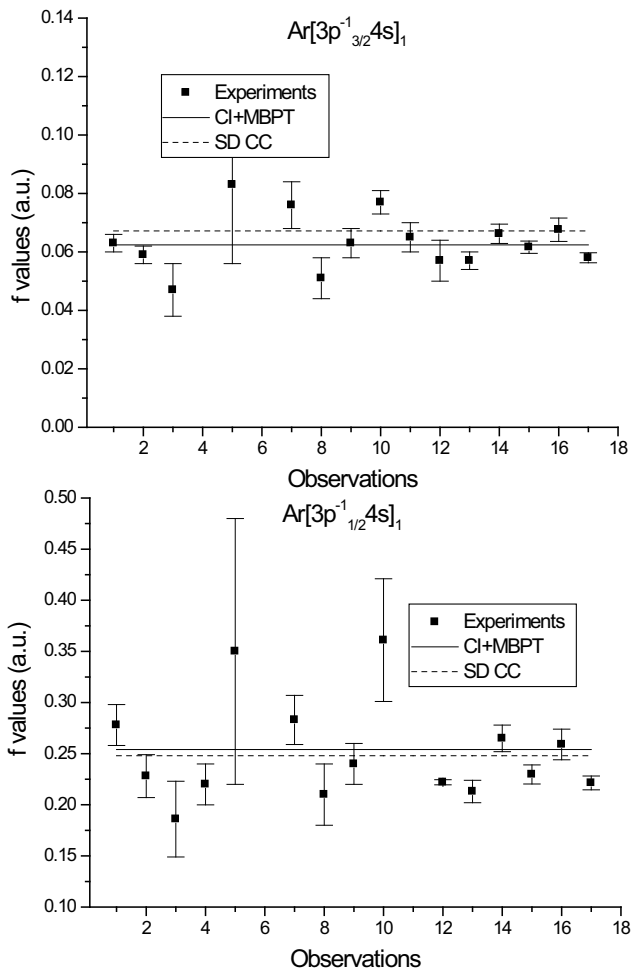}}
\caption{Comparison of CI+MBPT with experiment and SD CC
calculations~\cite{neon:98} for oscillator strengths of the
$[p_{3/2}^{-1}4s]_1$ and $[p_{1/2}^{-1}4s]_1$ states of argon}
\label{argexp1}
\end{figure}
\begin{table}[tbp]
\caption{References for experimental data shown in
Fig.~\ref{argexp1}} \label{graphtable}
\begin{center}
\begin{tabular}{lllcc}
\hline \hline
$\#$ & Reference & Year & $f$ 1s$_4$ & $f$ 1s$_2$ \\
\hline
1 & \citet{Lewis}       & 1967 & 0.063$\pm$0.003 & 0.278$\pm$0.020\\
2 & \citet{Lawrence2}    & 1968 & 0.059$\pm$0.003 & 0.228$\pm$0.021\\
3 & \citet{Geiger}      & 1970 & 0.047$\pm$0.009 & 0.186$\pm$0.037\\
4 & \citet{Jongh}       & 1971 &                 & 0.22$\pm$0.02\\
5 & \citet{Irwin2}       & 1973 & 0.083$\pm$0.027 & 0.35$\pm$0.13\\
6 & McConkey and        &       &                 &              \\
  & Donaldson~\cite{McConkey}    & 1973 & 0.096$\pm$0.002 &              \\
7 & \citet{Copley}      & 1974 & 0.076$\pm$0.008 & 0.283$\pm$0.024\\
8 & \citet{Vallee}      & 1977 & 0.051$\pm$0.007 & 0.210$\pm$0.030\\
9 & \citet{Westerveld}  & 1979 & 0.063$\pm$0.005 & 0.240$\pm$0.020\\
10 & Hahn and           &      &                 &                \\
  & Schwentner~\cite{Hahn}& 1980 & 0.077$\pm$0.004 & 0.361$\pm$0.060\\
11 & \citet{Chornay}    & 1984 & 0.065$\pm$0.005 &                \\
12 & \citet{Li}         & 1988 & 0.058$^{+0.005}_{-0.008}$ & 0.222$^{+0.002}_{-0.003}$\\
13 & \citet{Tsurubuchi} & 1990 & 0.057$\pm$0.003 & 0.213$\pm$0.011\\
14 & \citet{Chan}       & 1992 & 0.0662$\pm$0.0033 & 0.265$\pm$0.013\\
15 & \citet{Lightenberg}& 1994 & 0.0616$\pm$0.0021 & 0.2297$\pm$0.0093\\
14 & \citet{Wu}         & 1995 & 0.0676$\pm$0.0040 & 0.259$\pm$0.015\\
15 & \citet{gibbson}     & 1995 & 0.0580$\pm$0.0017 & 0.2214$\pm$0.0068\\
\hline
\end{tabular}
\end{center}
\end{table}
We calculate oscillator strengths for well-studied experimentally
and theoretically resonant argon transitions. The upper and lower
panels of Fig.~\ref{argexp1} show data for the
$[3p_{3/2}^{-1}4s]_1$ and $[3p_{1/2}^{-1}4s]_1$ states,
respectively. The agreement with SD CC calculations~\cite{neon:98}
and most experiments is very close. For the weaker transition,
deviation from the SD CC theory is somewhat larger, which is due
to cancellation of contributions from different configurations.
Such cancellation was more complete in neon, where we observed
larger disagreement with the theory of Ref.~\cite{neon:98}.

\subsection{Energies of argon even J=2 states}

Table~\ref{arg20} shows our calculated energies for J=2 even
states. Results for these and J=1 odd states are similar: fine
structure for 4p, 5p, or 4f states is well reproduced; a small
systematic shift exists due hole energy offset which if subtracted
brings theory in a better agreement for less correlated higher
excited states. The convergence with the number of particle-hole
configurations is achieved for the size of CI space 48.

\begin{table}[tbp]
\caption{Energies of argon even J=2 states. For unique
specification of levels, accurate NIST energies are provided. All
energies are expressed in atomic units. $\Delta_{7}$ is the
deviation from experiment for the 7th level} \label{arg20}
\begin{center}
\begin{tabular}{ccccccc}
\hline\hline State & NIST  & CI-6  & CI-24  & CI-48  & $\Delta$ &
$\Delta-\Delta_{7}$ \\ \hline
p$_{3/2}^{-1}$4p & 0.48123 & 0.4838 & 0.4837 & 0.4844 & 0.0032 & -0.0050 \\
p$_{3/2}^{-1}$4p & 0.48405 & 0.4870 & 0.4871 & 0.4878 & 0.0037 & -0.0045 \\
p$_{1/2}^{-1}$4p & 0.48885 & 0.4919 & 0.4920 & 0.4928 & 0.0039 & -0.0043 \\
p$_{3/2}^{-1}$5p & 0.53309 & 0.5397 & 0.5396 & 0.5403 & 0.0072 & -0.0010 \\
p$_{3/2}^{-1}$5p & 0.53393 & 0.5405 & 0.5406 & 0.5413 & 0.0074 & -0.0008 \\
p$_{1/2}^{-1}$5p & 0.53978 & 0.5466 & 0.5466 & 0.5474 & 0.0076 & -0.0006 \\
p$_{3/2}^{-1}$4f & 0.54762 & - & 0.5597 & 0.5562 & 0.0086 & 0.0004 \\
p$_{3/2}^{-1}$4f & 0.54781 & - & 0.5601 & 0.5564 & 0.0086 & 0.0004 \\
p$_{3/2}^{-1}$6p & 0.55219 & - & 0.5665 & 0.5604 & 0.0082 & 0.0000
\\ \hline
\end{tabular}
\end{center}
\end{table}
\subsection{Argon transitions between excited states}
\begin{table}[tbp]
\caption{Line strengths in atomic units for transitions between
J=1 odd and J=2 even states. Accuracy of experiments is shown as
letters in accordance with the NIST convention~\cite{nist:01}.
Three cases with respect to RPA corrections are considered:
``cRPA'' (conventional), ``nRPA'' (no), and ``mRPA''
(modified-denominator) RPA corrections are included.
The deviation from experiment is measured by $\Delta $, which is defined as $%
\Delta =(S_{\text{exp.}}-S_{\text{theory}})/S_{\text{exp.}}$}
\label{argls}
\begin{center}
{\small
\begin{tabular}{ccclrrrrrr}
\hline\hline Transition & $\lambda $,\AA & Exp. & Ac. & cRPA &
nRPA  & mRPA & $\Delta $ \\ \hline
p$_{3/2}^{-1}$4s-p$_{3/2}^{-1}$4p & 8427 & 31.8 & C & 24.7 & 31.6
& 27.8 & 13\% \\
p$_{3/2}^{-1}$4s-p$_{3/2}^{-1}$4p & 8008 & 6.21 & C & 2.60 &  3.28
 & 2.94 & 53\% \\
p$_{3/2}^{-1}$4s-p$_{1/2}^{-1}$4p & 7386 & 8.42 & C & 7.31 &  9.41
 & 8.23 & 2\% \\
p$_{3/2}^{-1}$4s-p$_{3/2}^{-1}$5p & 4301 & 0.074 & C & 0.016  &
0.15  & 0.062 & -16\% \\
p$_{3/2}^{-1}$4s-p$_{3/2}^{-1}$5p & 4267 & 0.060 & C & 0.025  &
0.079
 & 0.048 & 20\% \\
p$_{3/2}^{-1}$4s-p$_{1/2}^{-1}$5p & 4046 & 0.054 & C & 0.025  &
0.099
 & 0.054 & 0\% \\
p$_{1/2}^{-1}$4s-p$_{3/2}^{-1}$4p & 9787 & 3.40 & C & 3.41 & 4.26
 & 3.79 & -12\% \\
p$_{1/2}^{-1}$4s-p$_{3/2}^{-1}$4p & 9227 & 9.75 & C & 10.9 & 13.8
& 12.2 & -25\% \\
p$_{1/2}^{-1}$4s-p$_{1/2}^{-1}$4p & 8411 & 32.7 & C & 22.7 & 28.9
 & 25.6 & 22\% \\
p$_{1/2}^{-1}$4s-p$_{3/2}^{-1}$5p & 4630 & 0.0094 & C- & 0.0013  &
0.024  & 0.0097 & -3\% \\
p$_{1/2}^{-1}$4s-p$_{3/2}^{-1}$5p & 4591 & 0.0015 & D & 0.0003  &
0.0041  & 0.0006 & -60\% \\
p$_{1/2}^{-1}$4s-p$_{1/2}^{-1}$5p & 4335 & 0.114 & C & 0.030 &
0.214  & 0.0985 & 14\% \\ \hline
\end{tabular}
}
\end{center}
\end{table}

\begin{table}[tbp]
\caption{Comparison of transition rates  for transitions between
J=1 odd and J=2 even states. NIST values represent experimental
results, our transition rates are obtained from mRPA line strenths
in the previous table using experimental wavelengths. Wavelengths
specify transitions shown in the previous table. In addition to
experiment, our results are also compared with various
semiempirical calculations. Rates are expressed in $s^{-1}$.
Brackets denote powers of 10} \label{argrate}
\begin{center}
{\small
\begin{tabular}{crrrrrrr}
\hline\hline $\lambda $,\AA & NIST &mRPA& Ref.\cite{Lilly} &
Ref.\cite{gvb} &Ref.\cite{Wende}  & Ref.\cite{Bues}
& Ref.\cite{jones} \\
\hline
8427 & 2.15[7] & 1.88[7] & 2.08[7] & 2.11[7] & -& - & -\\
 8008 & 4.90[6] & 2.32[6] & 4.93[6] & 4.8[6] & - & - & -\\
 7386 & 8.47[6] & 8.28[6] & 9.18[6] & 9.11[6]& - & - & -\\
 4301 & 3.77[5] & 3.16[5] & 1.33[6]  &1.3[6]  & 3.77[5] & 3.77[5]& 3.77[5] \\
 4267 & 3.13[5] & 2.50[5] & 3.90[5] &4.6[5]& 2.80[5]& 3.51[5] & 3.05[5]\\
 4046 & 3.30[5] & 3.30[5] & 5.30[5] &5.2[5]& 3.40[5] & 3.10[5]& 3.49[5] \\
 9787 & 1.47[6] & 1.64[6] & 1.25[6] & 1.07[6]& - & - & -\\
 9227 & 5.03[6] & 6.29[6] & 5.24[6] & 5.9[6]& - & - & -\\
 8411 & 2.23[7] & 1.74[7] & 2.25[7] & 2.21[7]& - & -& -\\
 4630 & 3.84[4] & 3.96[4] & 2.20[5]  &2.0[5]  & 3.9[4] & 3.7[4]& 3.9[4] \\
 4591 & 6.28[3] & 2.51[3] & 1.50[5]  &1.6[5]  & 3.60[3] & 7[3] & 8[3]\\
 4335 & 5.67[5] & 4.90[5] & 1.66[6]  &1.7[6]  & 5.90[6] & 5.53[6]&5.62[6] \\
\hline
\end{tabular}
}
\end{center}
\end{table}

\begin{figure}[tbp] \centerline{\includegraphics*[scale=0.49]{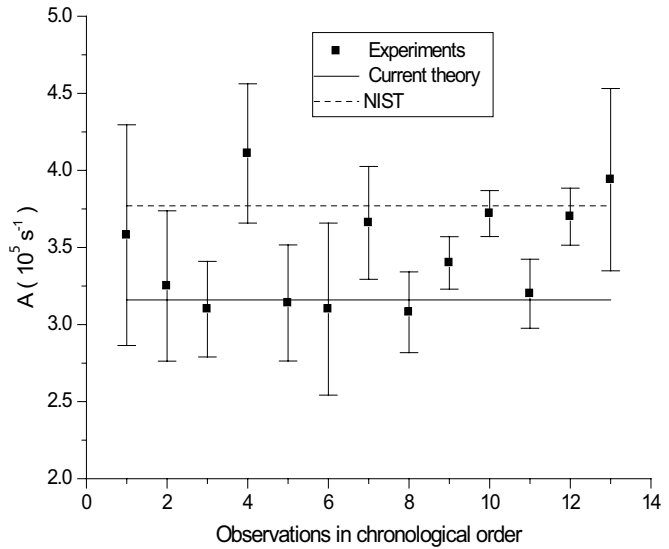}}
\caption{Comparison of CI-MBPT and experimental transition
probabilities for the 430-nm argon transition} \label{arg430}
\end{figure}
\begin{table}[tbp]
\caption{References for experimental data shown in
Fig.~\ref{arg430}} \label{graphtable}
\begin{center}
{\small
\begin{tabular}{llccc}
\hline \hline
Obs. & Reference & Year & $A$ (10$^5$ s$^{-1}$) & $\sigma$ \\
\hline
1 & \citet{Drawin}   & 1956 &3.58 & 20\% \\
2 & \citet{Gericke}  & 1961 &3.25 & 15\% \\
3 & \citet{Richter}  & 1965 &3.1  & 10\% \\
4 & \citet{Popenoe}  & 1965 &4.11 & 11\% \\
5 & \citet{Chapelle}& 1968 &3.14 & 12\% \\
6 & \citet{Wende}    &1968& 3.1& 18\% \\
7 & \citet{Wujec}    & 1971&3.66& 10\% \\
8 & \citet{Houwelingen} & 1971 &3.08 & 8.5\% \\
9 & \citet{Nubbemeyer}  & 1976 &3.40 &5\% \\
10 & \citet{Preston}    & 1977 &3.72 &4\% \\
11 & \citet{Baessler}   & 1980 & 3.2 & 7\% \\
12 & \citet{Hirabayashi}& 1987 & 3.94& 15\% \\
\hline
\end{tabular}
}
\end{center}
\end{table}
Table~\ref{argls} shows the comparison of our calculations with
experiment for line strengths between J=1 odd and J=2 even state
transitions. Very large disagreement is observed when no RPA
correction is added. However, if conventional  RPA correction
(Eq.\ref{RPA1})
 is included, the difference is also large, though of an
opposite sign. And only if we include the appropriate,
modified-denominator RPA correction (mRPA), which constitutes
roughly a half of the conventional RPA correction, do our values
agree with experiment at the level of the experimental accuracy.
The experimental accuracy in Table~\ref{argls} is denoted by
letters in accordance with the NIST convention: for example, the
accuracy of class ``C'' is about 10\%. Still in the case of the
p$_{3/2}^{-1}$4s-p$_{3/2}^{-1}$4p transition the disagreement is
quite large and independent of RPA corrections which might
indicate that the problem is in the experiment. If we exclude this
case and the inaccurate measurement of class ``D'' , then mRPA
calculations can be considered accurate at the level of 10-20\%.

We also compare our final results (mRPA calculations) with some
available semiempirical calculations by
\citet{Lilly,gvb,Wende,Bues,jones}. In some cases, semiempirical
results are close to experiment, but in others they can deviate
substantially, by orders of magnitude, and they also disagree with
each other. For example, for weak transitions, transition rates of
\citet{Lilly} are about 10 times larger than experimental values;
for these transitions results of \citet{gvb} are much better. Our
theory agrees with experiment more consistently in cases where
experiments have good precision.

For one particular transition  at 430 nm, many measurements have
been conducted, and this transition can be considered to be known
most accurately among other transitions between excited states of
argon. In Fig.~\ref{arg430} we compare our theoretical value with
many available experimental values. While six experiments support
NIST quoted value, a greater number of experiments support our
theoretical value. Therefore, we tend to concluded that NIST
values can be somewhat wrong. We also find that semiempirical
calculations are not able to predict these particular experiments;
namely, most extensive semiempirical calculations \cite{Lilly}
give the transition rate $1.33\times 10^{6}$ s$^{-1}$ that
disagrees with all experiments. Returning to our
Table~\ref{argls}, we can argue that some large deviations of our
theory from experiment can be caused by inaccuracy of quoted
experimental errors.

For better tests of our theory, more accurate experiments are
needed. For this reason, we have measured the ratio of line
strengths using an accurate laser absorption
technique~\cite{isavargexe}. Our experimental ratio,
$3.29\pm0.13$, for the two transitions at 9787 \AA ~and 9227 \AA
~completely agrees with our theoretical value, 3.22. It would be
extremely interesting if this experimental technique was further
developed and applied to other transitions in argon or in other
noble gas atoms. For example, experiments in neon using visible
lasers could test the accuracy of this experimental method, since
our theory is expected to be more accurate in neon than in argon.

\section{Conclusions}

In this paper, we have applied CI+MBPT theory  to calculate
energies and transition properties of argon particle-hole states.
Our theoretical energies and transition rates are in good
agreement with most experiments. We found that random-phase
approximation corrections are essential. We also analyzed
dependence of accuracy on the size of configuration space and
determined the optimal size of CI matrix. Our CI+MBPT program can
be used for calculations of many other transitions of closed-shell
atoms and ions.


\end{document}